\documentclass[a4paper]{jpconf}
\usepackage{graphicx}
\usepackage{hyperref}
\begin{document}
\title{The FUNK search for Hidden Photon Dark Matter in the eV range}

\author{Babette D\"obrich$^{1,\star}$, Kai Daumiller$^2$, Ralph Engel$^2$, Joerg Jaeckel$^3$,
Marek Kowalski$^{4,5}$, Axel Lindner$^4$, Hermann-Josef Mathes$^2$, Javier Redondo$^{6,7}$,
Markus Roth$^2$, Christoph Sch\"afer$^2$, Ralf Ulrich$^2$, Darko Veberic$^2$}

\address{$^1$ CERN, 1211 Geneva 23, Switzerland}
\address{$^2$ Institute for Nuclear Physics, Karlsruhe Institute of Technology (KIT), Germany}
\address{$^3$ Institute for Theoretical Physics, Heidelberg University, Germany}
\address{$^4$
Deutsches Elektronen-Synchrotron DESY, Hamburg and Zeuthen, Germany}
\address{$^5$
Department of Physics, Humboldt University, Berlin, Germany}
\address{$^6$
Department of Theoretical Physics, University of Zaragoza, Spain}
\address{$^7$
MPI f\"ur Physik, Munich, Germany}

\ead{$\star$ babette.dobrich@cern.ch}

\begin{abstract}
We give a brief update on the search for Hidden Photon Dark Matter with FUNK.
The experiment uses a large spherical mirror, which, if Hidden Photon Dark Matter exists in the
accessible  
mass and coupling parameter range, would yield an optical signal in the mirror's center in an otherwise
dark environment. After a test run with a CCD, preparations for a run with a low-noise PMT are under way
and described in this proceedings.
\end{abstract}

\section{Holding a mirror up to Hidden Photon Dark Matter}

There is compelling evidence for Dark Matter, except that we do not know what its constituents are. 
The possibility of  `heavier' Dark Matter is under investigation by a large number of direct detection experiments.
The possible signature of (possibly annihilating) Dark Matter is investigated in astrophysical settings or with
direct
production at colliders, e.g. at the LHC. For a recent review, see \cite{Klasen:2015uma}.

For very light-weight Dark Matter fewer experiments exist, see, e.g., \cite{Dobrich:2015xca}
for a brief review. Most prominently, a growing number of experiments   
are hunting after axionic Dark Matter \cite{patras}. Recently, it was suggested that also ultra-light extra `hidden'
gauge bosons make up viable Dark Matter, either through a misalignment mechanism \cite{Nelson:2011sf,Arias:2012az}
(similarly to axions), or sourced by
inflationary fluctuations \cite{Graham:2015rva}.

In the experiment described here, the relevant signature for the presence of Hidden Photon Dark Matter comes from
a possible
photon-to-hidden-photon coupling,
$\sim \chi F^{\mu \nu} X_{\mu \nu}$,
with visible and `hidden' electromagnetic field-strengths  $ F^{\mu \nu}$ and $ X^{\mu \nu}$, respectively.
Their coupling strength is parameterized by  the kinetic
mixing parameter $\chi$, see, e.g. \cite{Jaeckel:2013ija} for a review on this physics. 
Additionally, the Hidden Photon
should have a  mass $m$ through a Higgs or St\"uckelberg mechanism.

\begin{figure}[h]
\includegraphics[width=20pc]{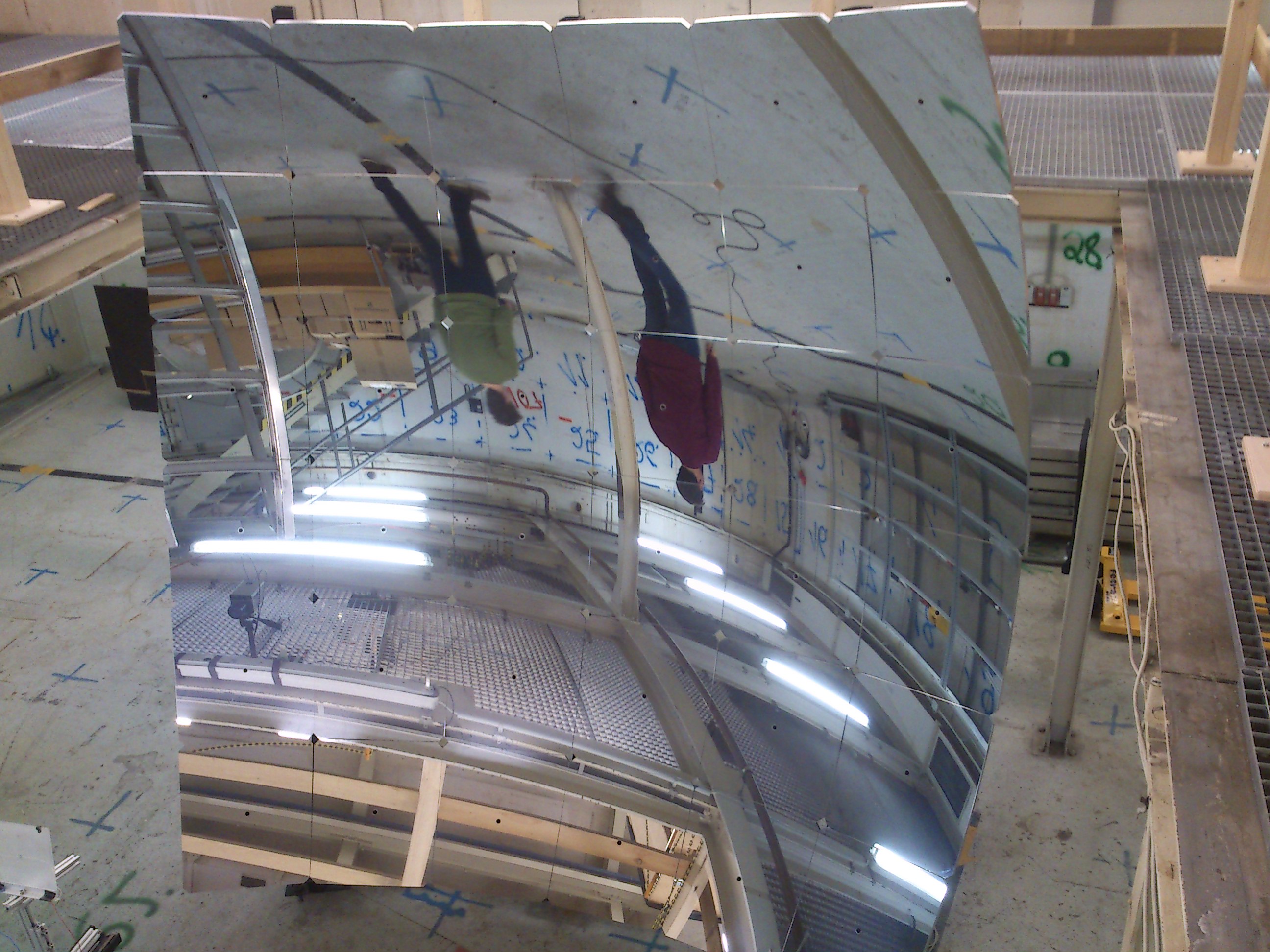}\hspace{2pc}%
\begin{minipage}[b]{14pc}\caption{\label{mirror}
Mirror used in the search of Hidden Photon Dark Matter with FUNK.
The mirror is set up in a window-less hall at Karlsruhe Institute
for Technology, Northern Campus.
On the left-hand side of the picture is the mount for the detector (in the picture it carries a frosted
glass used in the alignment of the mirror elements).}
\end{minipage}
\end{figure}

The FUNK (Finding U(1)s of a Novel Kind) experiment \cite{Dobrich:2014kda} is based on the technique described in \cite{Horns:2012jf}: Hidden Photon Dark Matter can induce radiation emitted off a metallic surface
due to its kinetic mixing with photons (which yields an effective coupling to
the electrons in the surface).
In our experiment, this mirror is composed of $6\times6$ mirror elements originally produced for
the fluorescence detector of the Pierre Auger Observatory. The total surface area of this `FUNK mirror' is about 
14\,m$^2$, see Fig.~\ref{mirror}.

Note that as Dark Matter, the Hidden Photons are almost at rest with respect to the mirror.
In effect, the electrons in the conducting mirror surface see a very small oscillating electric field (strength downscaled through $\chi$)
and the electrons can oscillate in phase and therefore emit a wave mostly {\it  perpendicular} to the mirror.
Thus, the signal is expected in the radius point $R$, not the focal point which makes 
measuring a potential signal considerably easier.
 A small angular offset from the rectangular emission  is determined by the relative velocity of the 
Hidden Photons with respect to the mirror.

Such an experiment is sensitive to all Hidden Photon masses 
whose associated wavelength $\lambda=2\pi/m$ can be: 
(1) reflected from the mirror (the induced emissivity is related to the reflectivity), (2) properly concentrated in the focal plane (wavelength longer than typical imperfections) and (3) detected by a detector in the center of the sphere

An experiment also using this  technique, albeit a smaller mirror (diameter of $50$ cm with
a focal length of $1$ m), is based in Tokio, and has recently published first results \cite{Suzuki:2015sza}
in the optical. Another dish configuration there is used to study the microwave regime \cite{Suzuki:2015vka}.

\section{Brief update on the `PMT run'}

The first measurements with FUNK  were done using a readily available `Sensicam' CCD whose characteristics are not sufficient to
reach un-tested Dark Matter parameter space.
The reason is the huge amount of pixels illuminated by a potential signal
and the correspondingly large dark noise contribution, see \cite{Veberic:2015yua} for results
and details.

For FUNK, due to careful alignment, the 90 \%
spot radius of the  mirror elements was reduced to $\sim$2\,mm as  described in \cite{Veberic:2015yua}.
The signal spot induced by Hidden Photon Dark Matter will also have a finite size due to the velocity distribution
of the Dark Matter $\Delta v=10^{-3}$ (in natural units).
Following \cite{Jaeckel:2013sqa}, the spot-radius for the DM-
induced spot will be 
$\Delta d \sim \Delta v R \sim 1 \ {\rm mm} \left(R/{\rm m} \right)=3.4$\,mm,
where now $R=3.4$ m denotes the radius of curvature of the mirror for FUNK.
The effective movement of the signal spot due the earth's movement with respect to the Dark Matter
is of a similar size.

 Our results of the test run with the CCD were reported  recently, 
 see \cite{Veberic:2015yua}, and we will therefore not
detail on them herein.

For the subsequent measurements in the optical, we decided to procure a 
photomultiplier tube (PMT) from ET Enterprise, model 9107QB \cite{QB}
with an active diameter of 25 mm 
to safely accommodate the full potential signal spot and its movement\footnote{
For the non-customized options, the other option `9893/350B' would have been 8mm
diameter which is too small in this setup.
%
 }.
The spectral coverage ranges from about 200-550\,nm at peak efficiencies of 28 \%.

The choice was mainly based on the favorable behavior of the multiplier series reported on in
\cite{Schwarz:2015lqa,schwarz,lozza} for the `9893/350B
type' with an active diameter\footnote{The PMT is originally sensitive at 52\,mm diameter but the
sensitive area is internally limited to 8\,mm diameter}  of 8\,mm.

This series of PMTs can be installed in
a `FACT50' housing that is designed to cool the PMT  up to $-50^\circ$C
beneath room temperature while regulating itself.
The behavior of the PMT in the cold is not specified by the vendor, 
but one reported Dark Count rate at $-20^\circ$C is $0.35 \pm 0.02$ Hz
in 30000s of acquisition time,  see \cite{lozza}.
With a customized blind-flange that can be attached to the FACT50 housing
or a `dark box'
the PMT can be tested under light-tight conditions.

\begin{figure}[h]
\includegraphics[width=\textwidth]{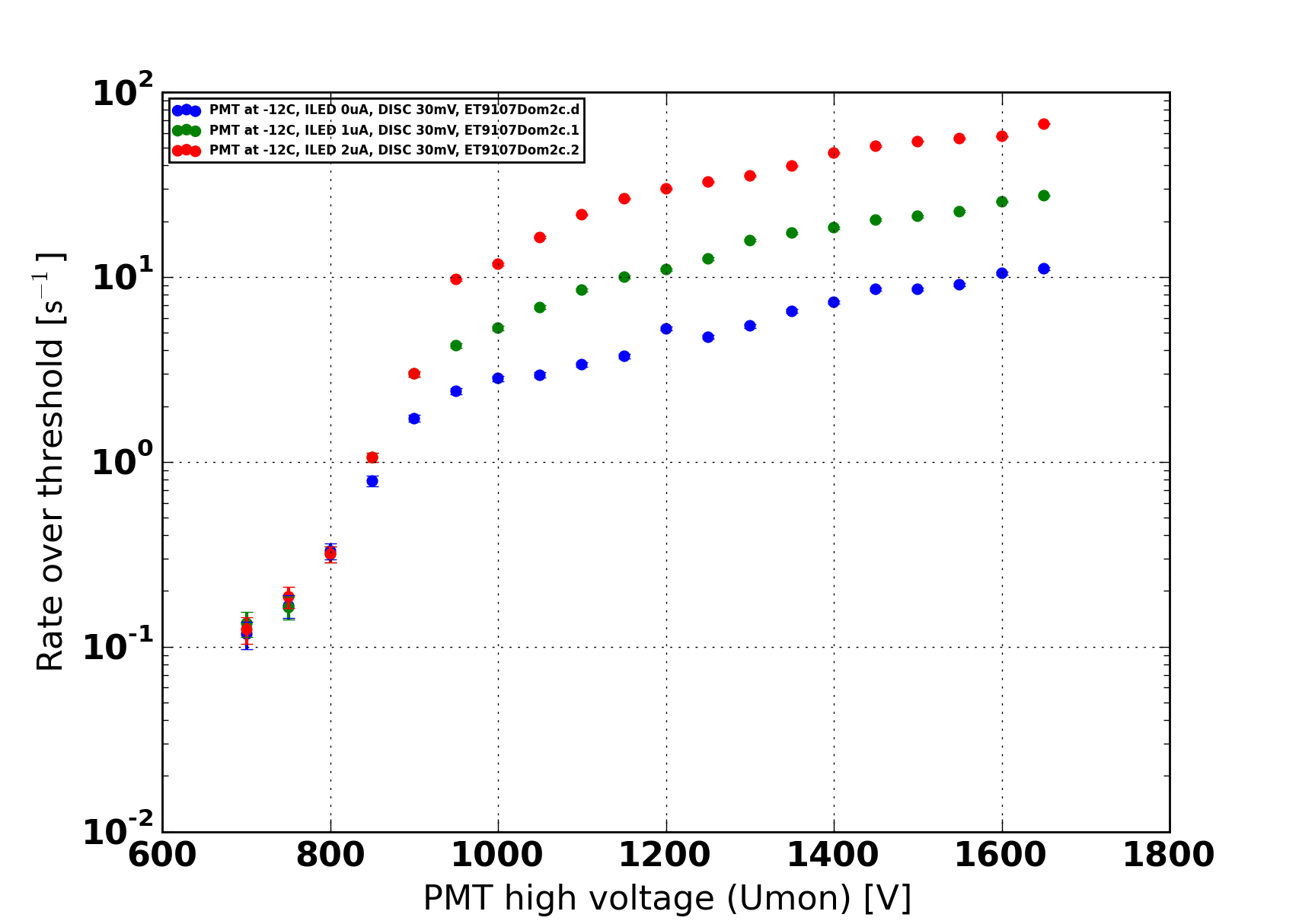}
\caption{\label{rates}
Rates over threshold for 9107QB in the FACT50 housing as function of applied high voltage
at a temperature of $-12^\circ$C. The three measurements correspond
to different light levels of an LED. Lowermost (blue curve) is  a `dark' environment,
middle (green) and lower (red) increasing LED currents.
  }
\end{figure}

In order to set the
most suitable working point for the photo-tube in single-photon counting
mode several characterization measurements need to be done. A crucial
feature of the individual photomultiplier is the so called plateau range
and especially its onset. Figure \ref{rates} shows the count rates at various
low light levels depending on the high voltage applied.

For a high voltage slightly above this onset of the plateau, the single
electron response of the PMT is currently being studied. This allows us to later
set lower and upper trigger thresholds in order to minimize further
non-single photon signals.

After the PMT  characterization
is finalized, we will mount it on a linear translation stage
in the mirror center that can shift the PMT laterally for 
signal and background measurements, respectively (see Figure \ref{mirror},
where in that picture in the radius point a milk glass is mounted).

\section{Summary}

After alignment and a test run of the FUNK experiment
(reported on in \cite{Veberic:2015yua}),
we are preparing for running with a low-noise PMT that
is expected to be sensitive to
un-charted Hidden Photon Dark Matter space in the eV region.

\ack
 B.D. thanks the organizers of PHOTON 2015 for a wonderful workshop in Academgorodok
and its participants for creating an enjoyable atmosphere.
The FUNK collaborators acknowledge support by the Helmholtz Alliance for Astroparticle Physics
and technical support by Hilmar Kr\"uger.

\end{document}